\pdfoutput=1
\documentclass[twocolumn,showpacs,twoside,10pt,pra]{revtex4}
\usepackage{graphicx}
\usepackage{subfigure}
\usepackage{epsfig}
\usepackage{epsf}
\usepackage{mathrsfs}
\usepackage{amssymb}
\usepackage{amsmath}
\usepackage{amsthm}
\usepackage{multirow}
\usepackage{hyperref}
\usepackage{graphicx}
\usepackage{hyperref}

\newcommand{\bra}[1]{\langle #1|}
\newcommand{\ket}[1]{|#1\rangle}
\usepackage[usenames]{color}
\definecolor{Red}{rgb}{1,0,0}
\definecolor{Blue}{rgb}{0,0,1}

\begin{document}

\title{Experimental realization of quantum algorithms for linear system inspired by adiabatic quantum computing}

\author{Jingwei Wen$^{1}$}
\thanks{These authors contributed equally to this work.}
\author{Xiangyu Kong$^{1}$}
\thanks{These authors contributed equally to this work.}
\author{Shijie Wei$^{4}$}
\author{Bixue Wang$^{1}$}
\author{Tao Xin$^{5}$}
\email{xint@sustc.edu.cn}
\author{Guilu Long $^{1,2,3}$}
\email{gllong@tsinghua.edu.cn}

\affiliation{$^{1}$State Key Laboratory of Low-Dimensional Quantum Physics and Department of Physics, Tsinghua University, Beijing 100084, China}

\affiliation{$^{2}$ Tsinghua National Laboratory for Information Science and
Technology,  Beijing 100084, P. R. China.}

\affiliation{$^{3}$  Collaborative Innovation Center of Quantum Matter, Beijing 100084, China}

\affiliation{$^{4}$  IBM research, Beijing 100094, China}

\affiliation{$^{5}$  Shenzhen Institute for Quantum Science and Engineering,
Southern University of Science and Technology, Shenzhen 518055, China}

\begin{abstract}
Quantum adiabatic algorithm is of vital importance in quantum computation field. It offers us an alternative approach to manipulate the system instead of quantum gate model. Recently, an interesting work [\href{https://arxiv.org/abs/1805.10549}{{\color{blue}arXiv:1805.10549}}] indicated that we can solve linear equation system via algorithm inspired by adiabatic quantum computing. Here we demonstrate the algorithm and realize the solution of 8-dimensional linear equations $A\textbf{x}=\textbf{b}$ in a 4-qubit nuclear magnetic resonance system. The result is by far the solution of maximum-dimensional linear equation with a limited number of qubits in experiments, which includes some ingenious simplifications. Our experiment provides the new possibility of solving so many practical problems related to linear equations systems and has the potential applications in designing the future quantum algorithms.

\end{abstract}


\maketitle

\section{Introduction}
Based on the principles of quantum mechanics, quantum computation presents a novel framework to design the efficient algorithms and boost the computation processing with respect to the classical situations. The research on quantum computing can not only bring us powerful ability to manipulate microscopic system, but also can provide us novel perspectives to understand the world and invent new techniques. Since the birth of quantum computation \cite{Feynman,benioff}, so many works has been done to apply the properties of quantum system to other research fields, such as communication \cite{qkd,qsdc1,qsdc2}, cryptography \cite{cry1,cry2}, and machine learning \cite{svm1,svm2}. Furthermore, many computation models have been put forward including circuit model \cite{cir1,cir2}, one-way quantum computing \cite{oneway}, topologic quantum computation \cite{topo1,topo2}, adiabatic quantum computing (AQC) \cite{aqc1,aqc2} and duality quantum computing \cite{dual}. Among them, AQC might be one of the most prospective models for practical application at the recent advances in quantum machine learning, because machine learning usually deals with a form of multivariate optimization, which can be directly translated to AQC\cite{aqc}.

In general, AQC starts with a time-dependent initial Hamiltonian $H_{0}$ which is conveniently prepared in experiments. By driving the initial Hamiltonian to the target Hamiltonian $H_{p}$, we could get the information encoded in the ground state of $H_{p}$ \cite{aqc1}. The transition from $H_{0}$ to $H_{p}$ is realized by driving an instantaneous Hamiltonian 
\begin{equation}
H(t)=[1-s(t)]H_{0}+s(t)H_{P}\, ,\label{e1}
\end{equation}
where the function $s(t)$ varies from 0 to 1 to parametrize the interpolation. Adiabatic theorem tells us that quantum systems tend to stay in the ground state of the instant Hamiltonian as long as the whole transformation process is sufficiently slow and smooth. Recently, some methods have been proposed to optimize the AQC process including randomization method (RM) \cite{RM} and shortcut to adiabatic passage (STAP) \cite{STAP}. 

The fundamental algorithm related to quantum machine learning was first proposed by Aram W. Harrow  \emph{et al} (HHL algorithm) \cite{hhl}. HHL algorithm is devoted to preparing a quantum state $\ket{x}$ representing the solution of a linear system of equation $A\textbf{x}=\textbf{b}$. Supposing that $A$ is a $N\times N$ matrix and $\textbf{b}$ is a $N$-dimensional vector, the best classical algorithm can find the solution with complexity in O(N), while the complexity of quantum HHL algorithm is polynomial in logN and $\kappa$, where $\kappa$ is called condition number, a parameter measuring the numerical instability of $A$. Recently, it has been shown that HHL algorithm can be neatly recast in the duality quantum computing formalism where linear combinations of unitary operators are used for computing \cite{dual, dual2}. Experimental realization of HHL algorithm has been demonstrated in nuclear magnetic system (NMR) \cite{exp2}, optical system \cite{exp3,exp4} and superconducting system \cite{exp5}. However, these experimental protocols only demonstrated the simplest situation of HHL algorithm by solving a $2 \times 2$ linear equation.

Both HHL algorithm and its experimental realization are based on gate model. Recently, an interesting approach has been proposed to implement algorithms inspired by adiabatic quantum computing for solving linear equations. In this work, we experimentally demonstrate two kinds of algorithms proposed in Ref \cite{aqchhl} using NMR system. On near-term quantum devices,  quantum resources will remain scarce and expensive. Thus, this approach shows a significant advantage on the consumption of qubit resources compared with HHL algorithm. 
This paper is organized as follows: In Sec \uppercase\expandafter{\romannumeral2}, we briefly review the basic theory. In Sec \uppercase\expandafter{\romannumeral3}, experimental setups and experimental procedure will be introduced. Then, we present the experimental results. Finally, a conclusion is given in Sec \uppercase\expandafter{\romannumeral4}.

\section{A brief review of the theory}
In this section, we briefly preview the basic framework of quantum algorithms introduced in Ref \cite{aqchhl}. The first quantum algorithm aims at changing system Hamiltonian from initial to the final form smoothly, keeping the quantum state staying at the ground state of instantaneous Hamiltonian
\begin{equation}
H(s)=A^{2}(s)-A(s)\ket{\overline{b}}\bra{\overline{b}}A(s)\, ,\label{e2}
\end{equation}
where $A(s)=(1-s)Z\otimes I+sX\otimes A$ and $\ket{\overline{b}}=\ket{+,b}$. The notations $I,X,Y,Z$ represent identity and Pauli matrix, and state $\ket{\pm}$ is the eigenstate of Hadamard gate in computational basis. The lower spectral gap bound between zero energy eigenstate and excited eigenstate energy of $H(s)$ is determined by parameter $s$ $(s\in[0,1])$ : $\Delta^{*}(s)=(1-s)^{2}+(s/\kappa)^{2}$  and $\kappa$ is condition number of $A$ matrix. Under the natural parametrization, $s(v)$ can be written as
\begin{equation}
s(v)=\frac{e^{v\frac{\sqrt{1+\kappa^{2}}}{\sqrt{2}\kappa}}+2\kappa^{2}-\kappa^{2}e^{-v\frac{\sqrt{1+\kappa^{2}}}{\sqrt{2}\kappa}}}{2(1+\kappa^{2})}\, ,\label{e3}
\end{equation}
where $v_{a} \leqslant v \leqslant v_{b}$ satisfying
\begin{equation}
v_{a}=\frac{\sqrt{2}\kappa}{\sqrt{1+\kappa^{2}}}log(\kappa \sqrt{1+\kappa^{2}}-\kappa^{2}),\label{e4}
\end{equation}
\begin{equation}
v_{b}=\frac{\sqrt{2}\kappa}{\sqrt{1+\kappa^{2}}}log(\sqrt{1+\kappa^{2}}+1).\label{e5}
\end{equation}

When parameter $v$ varies from boundary-value $v_{a}$ to $v_{b}$, $s(v)$ will increase progressively from $0$ to $1$. In this procedure, the eigenstate will correspondingly evolve from $\ket{-,b}$ to $\ket{+,x}$, and the target solution state $\ket{x}$ can be obtained by discarding the ancillary qubit. We choose fixed value $H(v^{j})$ of Hamiltonian in the $j$th step and evolve for a random time $t^{j}$ with $t^{j} \in[0~,~2\pi/\Delta^{*}(v^{j})]$, which is actually RM algorithm introduced in Ref\cite{RM}.

The second algorithm realizes energy gap amplification and improves the time complexity. In this algorithm, the system Hamiltonian is given by
\begin{equation}
H^{'}(s)=\sigma^{+}\otimes A(s)P^{\perp}_{\overline{b}}+\sigma^{-}P^{\perp}_{\overline{b}}A(s)\, ,\label{e6}
\end{equation}
where $\sigma^{\pm}=(X\pm iY)/2$ and $P^{\perp}_{\overline{b}}=I-\ket{\overline{b}}\bra{\overline{b}}$ is an orthogonal projector. The eigenvalues of $H^{'}(s)$ are $[~0,0,\pm\sqrt{\gamma_{1}(s)},...,\pm\sqrt{\gamma_{2N-1}(s)}~]$, where $\gamma_{j}(s)>0$ are the nonzero eigenvalues of $H(s)$. When we evolve the system from initial state $\ket{0}\otimes\ket{-,b}$, a series of projective measurement on the eigenstates of $H^{'}(s)$ will make the system end up staying in state $\ket{0}\otimes\ket{+,x}$ with sufficiently high probability. The fixed points we choose can be as same as the ones in the first algorithm, and we also evolve the Hamiltonian $H^{'}(v^{j})$ for a random time $t^{j}$ with $t^{j} \in[0~,~2\pi/\sqrt{\Delta^{*}(v^{j})}]$ at the $j$th step.

Compared with the first algorithm, the second algorithm introduces one more ancillary qubit, while the time complexity is decreased from $O(\kappa^{2}\text{log}(\kappa)/\epsilon)$ to $O(\kappa \text{log}(\kappa)/\epsilon)$, where $\epsilon \in (0,1)$ is  a precision parameter. Without phase estimation procedure, the number of ancillary qubits is independent with the size of quantum system. Thus, the expansibility of the algorithms can have a much better performance compared with HHL algorithm in spatial complexity.

\section{Experimental Setups and Results}

As proof-of-principle demonstrations, we experimentally realize the above two algorithms by solving an 8- and 4-dimensional linear equation, respectively. Because of the completeness of Pauli basis, we can expand matrix $A$ in an eight-dimension Hilbert space. Without the loss of generality, the linear equation we demonstrate in the first algorithm is $A\textbf{x}=\textbf{b}$, where matrix $A=(3III+XII-2XYI+3XYZ)/4$ and $\textbf{b}=[1,1,1,1,1,1,1,1]^{T}/\sqrt{8}$. It is worth emphasizing that the determination of matrix $A$ and vector $\textbf{b}$ is arbitrary, because this algorithm does not include any subroutines such as phase estimation which has been used in HHL algorithm. 

\begin{figure}
 	 \centering
 	  \subfigure[Molecule Structure]{
 	   \label{fig:subfig:a} 
  	   \includegraphics[width=4cm,height=3cm]{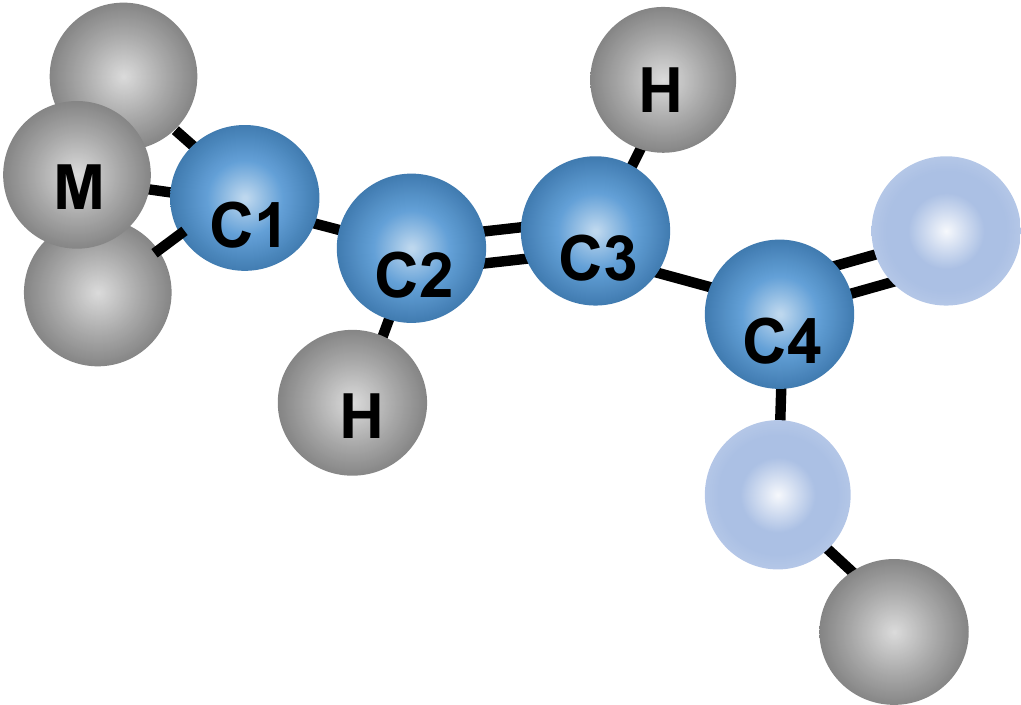}}
 	 \subfigure[Molecule Parameters]{
          \label{fig:subfig:b} 
  	  \includegraphics[width=4cm,height=3cm]{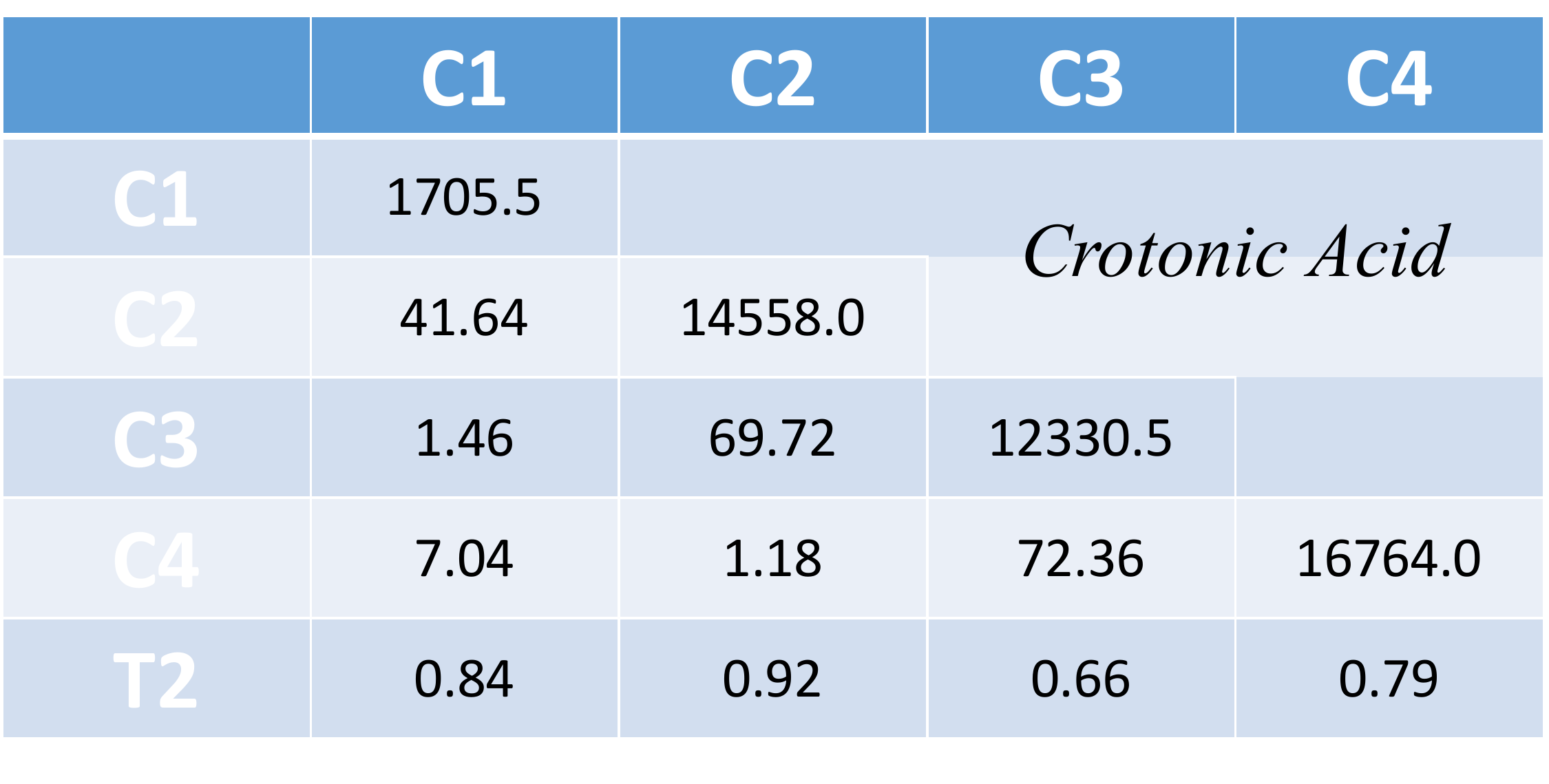}}
          \caption{(a) Molecule structure of ${}^{13}$C-labeled crotonic-acid. (b) Molecule parameters of sample: the chemical shifts and J-couplings (in Hz) are listed by the diagonal and off-diagonal elements, respectively. T2 (in Seconds) are also shown at bottom.}
          \label{fig:molecule} 
\end{figure}

In experiments, the used four-qubit sample is ${}^{13} C$-labeled transcrotonic acid dissolved in d6-acetone with $H$ decoupled throughout all the process. The structure and parameters of this molecule are shown in figure \ref{fig:molecule}. Notations C1 to C4 denote the four qubits, and we choose C1 as the ancillary qubit. The internal Hamiltonian under weak coupling approximation is
\begin{equation}
H=-\sum^{4}_{i=1}\pi\nu_{i}\sigma^{i}_{z}+\sum^{4}_{i<j}\frac{\pi}{2}J_{ij}\sigma^{i}_{z}\sigma^{j}_{z}\, ,\label{e8}
\end{equation}
where $\nu_{i}$ is the chemical shift and $J_{ij}$ is the J-coupling strength between the $i$th and $j$th nuclei. All experiments are carried out on a Bruker DRX 400MHz spectrometer at room temperature (296.5K).  

We start from the thermal equilibrium state and drive the system to the pseudo-pure state (PPS) using spatial averaging technique method\cite{pps_space1,pps_space2,pps_space3}. The procedure is realized by gradient fields and four unitary operators which is realized by Gradient Ascent Pulse Engineering(GRAPE) \cite{grape1,grape2} with the fidelity over $99.5\%$. The final form of four-qubit PPS is 
\begin{equation}
\ket{\rho_{0000}} = \frac{1-\epsilon}{16}I_{16} + \epsilon\ket{0000}\bra{0000}\, ,\label{e9}
\end{equation}
Where $I_{16}$ represents a $16\times 16$ identity operator and $\epsilon\approx10^{-5}$ is polarization. We first apply one $X$ gate on ancillary qubit followed by four Hadamard gates acting on all qubits, then we finish the preparation of the initial ground state $\ket{-,\textbf{b}}$.
As we introduced above, the evolution we want to realize is to slowly drive an instant Hamiltonian $H(t)$ which can be equivalently expressed by an unitary evolution $U\ket{-,\textbf{b}}=\ket{+,\textbf{x}}$. According to the RM theory mentioned above, we divide the total procedure into 300 steps: 
\begin{equation}
U=\prod_{i=1}^{300}U_{i}=U_{300} ...U_{i}...U_{2}U_{1},  
\end{equation}
where $U_{i}=e^{-iH(t_{i})\Delta_{i}}$. Notation $\Delta_{i}$ is the random evolution time of the $i$th step and $\Delta_{i} \in[0~,~2\pi/\Delta^{*}]$. In our experiment, we pack every 60 steps in one pulse which is also optimized by GRAPE method, with the length of each pulse 20ms and the fidelity with theoretical operators over 99\%.
In the end of circuit, we obtain the density matrix of final quantum state by performing quantum state tomography (QST) \cite{tomo1,tomo2,tomo3,tomo4,tomo5}. QST is finished by applying 17 readout pulses with the duration 0.9ms after the evolution. Then we can reconstruct all the density matrix elements of the final state $\rho_{f}$.

We perform four-qubit QST after each step and monitor the energy of system by using definition $\left \langle H \right \rangle=\text{tr}(\rho H)$. Experimental results are shown in Fig. \ref{fig:subfig:a}. It is shown that the ground energy of the system does not exceed the energy of first excited state within the range of experimental error, which means the process we realized is definitely adiabatic. Using the definition $F(\rho,\sigma)=\text{tr}(\rho\sigma)/\sqrt{\text{tr}\rho^2}\sqrt{\text{tr}\sigma^2}$\cite{fidelity}, the fidelity between the theoretical and experimental measured quantum state is over 95.5\% in the whole process of the experiment.
After tracing out the ancillary qubit, we find that the fidelity between experimental 3-qubit quantum state $\rho_{x}$ and theoretical solution $\ket{\textbf{x}}\bra{\textbf{x}}$ is about 98.4\%.
We reconstruct the quantum state from final density matrix and find the solution $x_{exp}$ in experiment (theoretical solution $x_{th}$), which is also shown in Fig. \ref{fig:subfig:b}. We also label the difference values between the experimental and theoretical data beside the bars in the picture.

\begin{figure}
 	 \centering
 	  \subfigure[ Energy Spectrum and Fidelity ]{
 	   \label{fig:subfig:a} 
  	   \includegraphics[width=8.5cm,height=5.5cm]{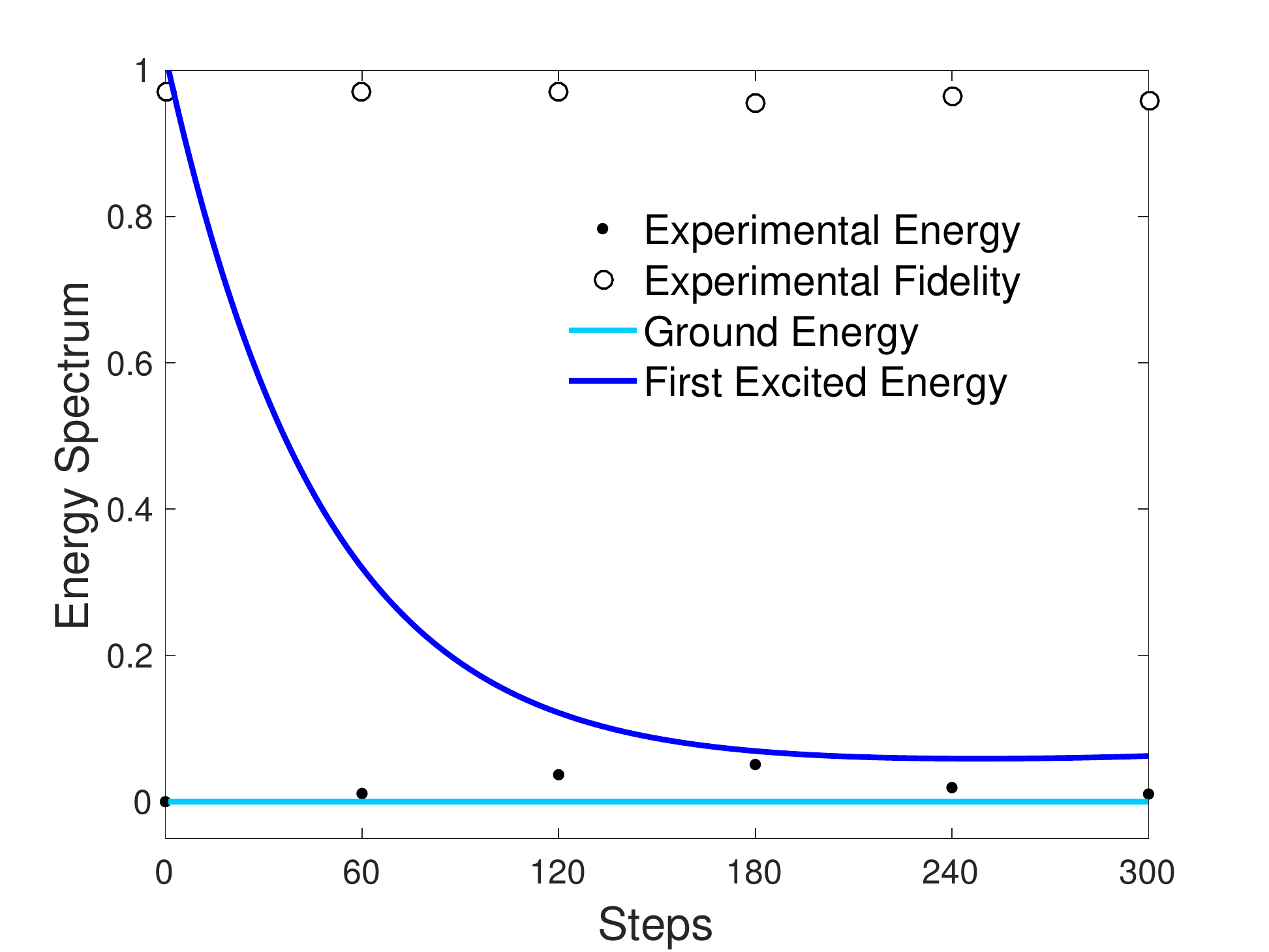}}
	\centering
 	 \subfigure[ Final Solution]{
          \label{fig:subfig:b} 
  	    \includegraphics[width=8.5cm,height=7.5cm]{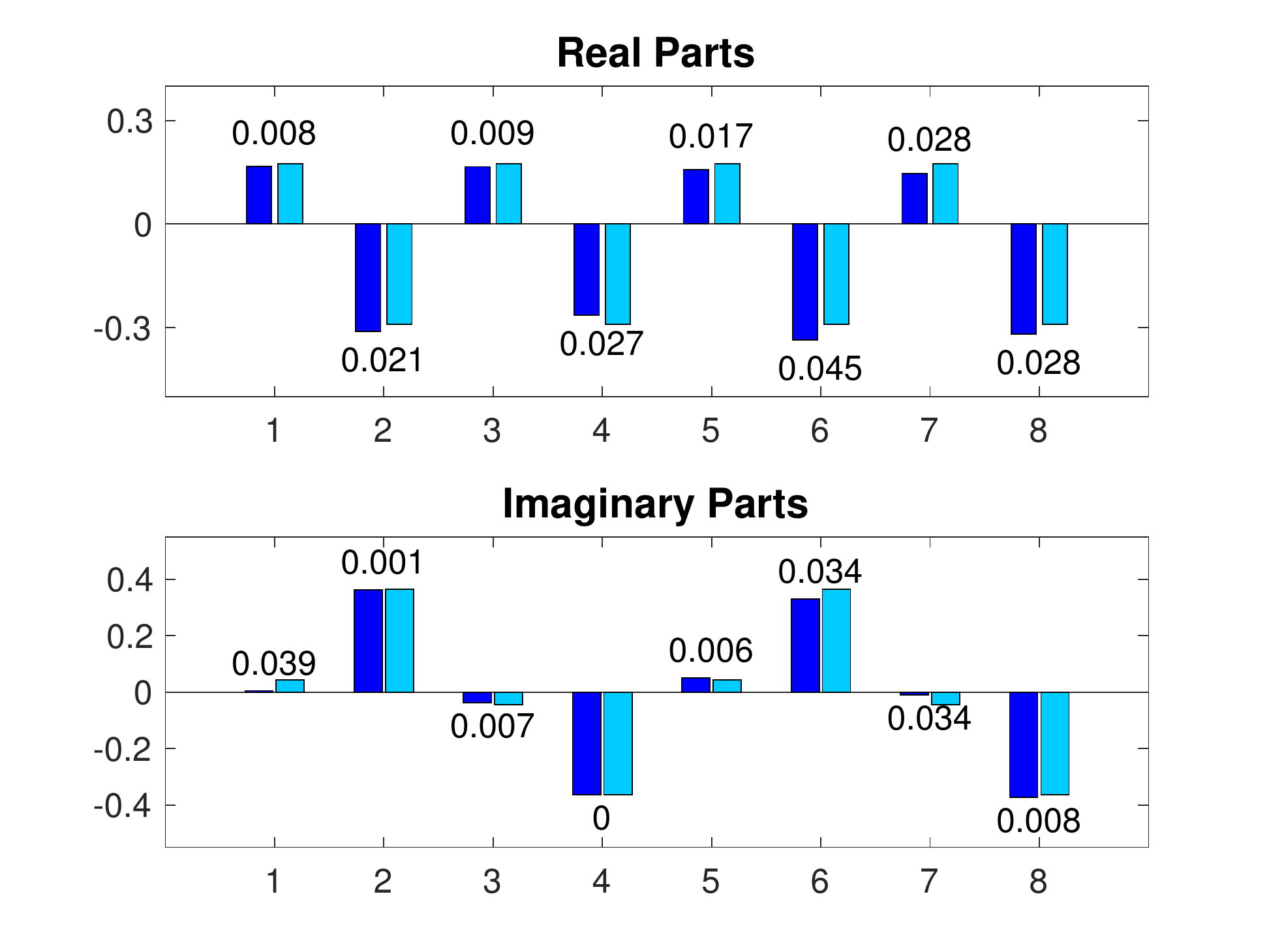}}
          \caption{Experimental results for the first algorithm. (a) The shallow and dark blue solid lines are theoretical values of the first-two energy level of the time-dependent Hamiltonian respectively.  Solid points (black below) represent experimental energy results and corresponding experimental fidelities are shown by circle points above. (b) The real and imaginary parts of the theoretical (shallow blue bars) and experimental (dark blue bars) final quantum state of solution $\ket{\textbf{x}}$ are shown. The numbers labeled present the corresponding differences between experimental and theoretical values.}
          \label{fig:results1} 
\end{figure}

By far, we have demonstrated the experimental realization of the first algorithm. In the following we would turn to the discussion of the second algorithm and its experimental results. In this case, we attempt to solve a 4-dimensional linear equation, where matrix $A$ and vector $\textbf{b}$ are chosen as $A=(3II+2ZI+3XI-3XY)/4$ and $\textbf{b}=[1,1,1,1]^{T}/2$. 

\begin{figure}
 	 \centering
 	  \subfigure{
  	   \includegraphics[width=8.5cm,height=7cm]{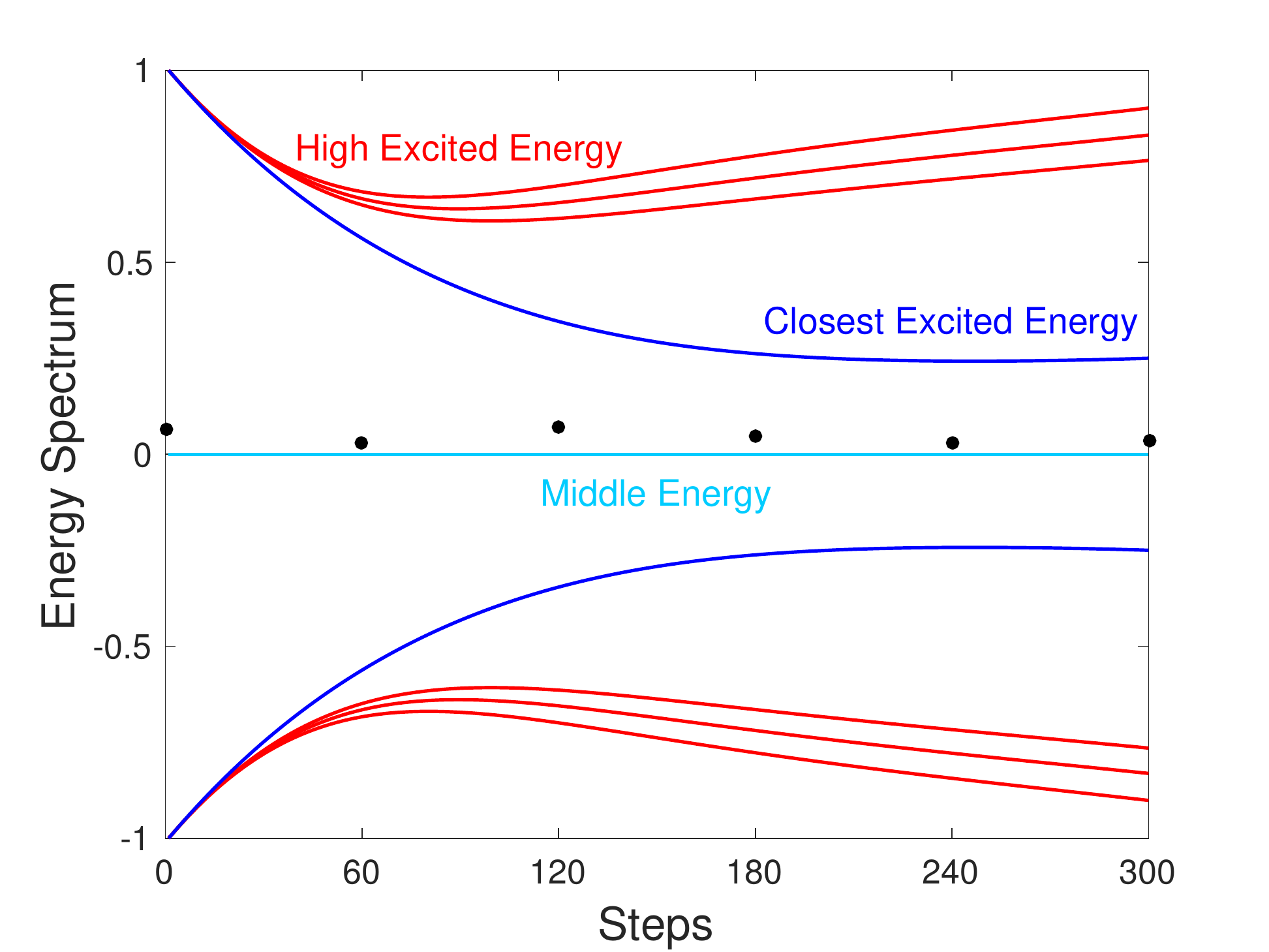}}
          \caption{Experimental results for the second algorithm. The solid lines are theoretical values of the energy spectrum of the time-dependent Hamiltonian. The shallow blue line is the middle level of the spectrum and the other lines represent the closest eight energy levels. Black points represent experimental energy results in the five steps of our experiment.}
          \label{fig:results2} 
\end{figure}

The experimental setups are almost the same: total numbers of the evolution steps we set are 300 and we also pack every 60 steps in one package and optimize them with GRAPE method. The four-qubit fidelities in experiments is all over 97\%. After tracing out the ancillary qubits (C1 and C2), the 2-qubit density matrix representing the solution is created and the 2-qubit fidelities is about 97.65\% in average. As a result, the experimental solution $x_{\text{exp}}=[0.157-0.039i, 0.193+0.009i, 0.454-0.590i, 0.509+0.352i]$, and the theoretical solution is $x_{\text{th}}=[0.175-0.019i, 0.175+0.019i, 0.500-0.468i, 0.500+0.468i]$.
We also list the energy levels of experimental and theoretical states in Fig. \ref{fig:results2}. The results indicate that the second algorithm does amplify the energy gap and the energy drift is merely 5\% of the energy gap between the ground energy level and the closest excited energy level. Therefore, the second algorithm obtains better adiabatic performance than the first one with identical evolution time, which means the second algorithm is more robust in AQC model.

\section{Conclusion}
In summary, we first demonstrate the solution of an $8$-dimensional linear equation system by utilizing algorithms inspired by adiabatic quantum computing in our NMR platform. The experimental results match well with theoretical expectations and we also compare the performance of the two different algorithms. To our knowledge, this is by far the solution of largest-dimensional linear equations realized in a quantum simulator. It is worth emphasizing that these algorithms have better expansibility than the previous algorithms. The determinations of matrix $A$ and vector $\textbf{b}$ are arbitrary and the complicated subroutines such as phase estimation and variable-time amplitude amplification are not necessary. In the future, one of the predominant challenges is how to establish a scalable physical system for quantum computation and resources of qubits are still very precious in the present development stage of quantum information. Under such a background, the protocols we demonstrate would be scalable and meaningful because the number of required ancillary 
qubits is independent on the size of quantum system. In experiments, we realize the demonstration of these algorithms by solving 8- and 4-dimensional linear equation with high fidelities. It is believed that the process we demonstrated can be extended to other quantum computing platforms.

\section{Acknowledgement}
This work was supported by the National Basic Research Program of China under Grant Nos. ~2017YFA0303700 and~2015CB921001, National Natural Science Foundation of China under Grant Nos. ~61726801,~11474168 and~11474181.

\end{document}